\documentclass[graybox, envcountchap]{svmult}

\usepackage{mathptmx}        % selects Times Roman as basic font
\usepackage{amsmath}
\usepackage{amssymb}
\usepackage{color}
\usepackage{helvet}          % selects Helvetica as sans-serif font
\usepackage{courier}         % selects Courier as typewriter font
\usepackage{dirtree}
\usepackage{makeidx}        % allows index generation
\usepackage{graphicx}        % standard LaTeX graphics tool
\usepackage{subfig}

\usepackage{multicol}        % used for the two-column index
\usepackage[bottom]{footmisc}% places footnotes at page bottom

\usepackage{hyperref}        %for hyperlinks
\hypersetup{colorlinks=true,urlcolor=blue}

\usepackage[misc]{ifsym}

\makeindex             % used for the subject index
\newcommand{\cala}{{\cal A}}
\newcommand{\calc}{{\cal C}}
\newcommand{\calo}{{\cal O}}
\newcommand{\calh}{{\cal H}}
\newcommand{\calr}{{\cal R}}
\newcommand{\calt}{{\cal T}}
\newcommand{\beq}{\begin{equation}}
\newcommand{\eeq}{\end{equation}}
\newcommand{\bea}{\begin{eqnarray}}
\newcommand{\eea}{\end{eqnarray}}
\newcommand{\hf}{\frac{1}{2}}

\begin{document}

%%%%%%%%%%%%%%%%%%%%%%%%%%%%%%%%%%%%%%%%%%%%%%%%%%%%%%%%%%%%%%%%%

\title{The unitarity crisis, nonviolent unitarization, and implications for quantum spacetime}
\author{Steven B. Giddings}
\institute{ (\Letter) \at Department of Physics, University of California, Santa Barbara, CA 93106 \email{giddings@ucsb.edu}}
\maketitle

\abstract{This contribution overviews the information paradox, or perhaps more aptly ``unitarity crisis," and a proposed resolution called nonviolent unitarization.
It begins by examining  the conflict of principles that yields the crisis, which can be phrased in terms of a ``black hole theorem" summarizing how basic assumptions come into conflict.
Proposed resolutions of the conflict, along with  problems with them, are overviewed.  The very important underlying question of localization of information and its role is discussed at some length, taking into account effects of perturbative gravity. The difficulty in finding a consistent scenario for black hole evolution strongly suggests new interactions on event horizon scales; a ``minimal" set of assumptions about these are parameterized in nonviolent unitarization.  Possible criticisms of this scenario, and some responses, are given.  New interactions at event horizon scales potentially lead to observable effects, via gravitational wave or electromagnetic channels, which are briefly discussed.  A possible origin of nonviolent unitarization effects from a more fundamental description of quantum spacetime, and possible implications for such a description, are also briefly discussed.
}

%%%%%%%%%%%%%%%%%%%%%%%%%%%%%%%%%%%%%%%%%%%%%%%%%%%%%%%%%

\section{Introduction}
\label{intro}

The black hole information paradox has evolved from what to many probably seemed an esoteric curiosity to what now appears to be a full blown crisis challenging the current foundations of physics.  In short, this problem appears to reveal an essential inconsistency between the basic principles of quantum mechanics, the principle of locality, and the principles of relativity, particularly the symmetries of Poincar\'e or general coordinate invariance.  These are the principles that provide the foundation of local quantum field theory -- which so far has beautifully described all natural phenomena except gravity -- and general relativity.  As will be discussed further, this is quite likely a crisis in the sense of Thomas Kuhn\cite{kuhn}, pointing toward the need for fundamental revision of current principles.  Since an essential question is that of unitary evolution, this is appropriately described as the {\it unitarity crisis}.

There have been many proposed resolutions to this crisis.  Typically these involve introduction of some extra structure, and then additional questions and problems arise; none have given a convincing resolution.  Frequently in the history of physics a conservative or minimalist approach has been successful, seeking a minimal revision to existing principles without introducing extraneous structure; many such revisions have proved to be profoundly radical.\footnote{Examples include Newtonian gravity, Special Relativity, Quantum Mechanics, General Relativity, and Quantum Field Theory.}  This contribution will explore such a conservative approach,  starting with some basic observations.  The first is that there is increasingly good evidence that black hole-like objects are ubiquitous in the cosmos, and so far appear to be well-described by classical general relativity.  Moreover, quantum mechanics has not only been tested in a wide range of contexts, but it has also been found to be very difficult to modify its basic principles without encountering extremely problematic behavior.  Finally, quantum field theory has proven to be an extremely accurate description of non-gravitational physics, in contexts with weakly-curved backgrounds.  Combining these with a careful formulation of the problem and  assuming that evolution is ultimately unitary suggests that a quantum black hole must have interactions with its surroundings that are not well-described by quantum field theory plus perturbatively-quantized general relativity, but that can be in a certain sense a small correction to that physics, and that ultimately unitarize evolution.  In particular, the failure so far to observe effects of structure near black hole horizons, and extrapolation of the statement that quantum field theory provides a highly-accurate description in weakly-curved regions, suggests that these interactions don't have violent effects on observers that might fall into a big black hole.  These are basic tenets of {\it nonviolent unitarization}.

Necessary background for reading this piece is a basic understanding of general relativity and of standard quantum field theory including techniques used in a traditional derivation of Hawking radiation\cite{Hawkrad}.  In outline, the next section will give a description of the essential conflict, formulated as a ``black hole theorem."  Section three will then give a more detailed description of aspects of the Hawking process, extending beyond the traditional analysis to a description of the evolving quantum state, and describing aspects of perturbative gravitational corrections, providing additional evidence that subtleties in such a description don't resolve the crisis, and providing a starting point for parameterizing corrections to such a standard description.  This initial discussion is important for explaining background motivation, in the spirit of the famous A.C. Doyle quote, through the words of Sherlock Holmes: ``How often have I said to you that when you have eliminated the impossible, whatever remains, however improbable, must be the truth?"
Section four then elaborates on the basic tenets of nonviolent unitarization, and section five describes its parametrization, and constraints on this, as a departure from the standard description.  Section six discusses some of the criticisms that have appeared in the literature or discussions, and responds to them.  Section seven introduces prospects for observational signatures, given our new era of multichannel observation of black holes.  Section eight concludes by discussing the question of connecting to a more foundational picture of quantum gravity, and the possibly quite radical ultimate implications of such a conservative approach.

\section{A black hole theorem and the crisis}
\label{bhthm}

\subsection{The black hole theorem}
\label{bhthms}

The unitarity crisis deals with the essential question of localization and transport of quantum information in quantum gravity.  In finite quantum systems, or in locally-finite ones such as lattice approximations to field theory, localization is described in terms of a factorization of the Hilbert space, {\it e.g.} $\calh=\calh_A\otimes \calh_B$.  With such a structure, we can 
consider independent excitations of subsystems $A$ or $B$, {\it e.g.} corresponding to localized information;  transfer of information can then be described by a hamiltonian that couples excitations in $A$ to those in $B$.

Before including gravity, local quantum field theory (LQFT) has a similar structure.  We may think of $A$ and $B$ as corresponding to spacelike-separated regions in spacetime.  However, 
there are additional subtleties associated to the infinite degrees of freedom of an unregulated theory.  Specifically, if we consider a region $U$ of a spacelike slice and its complement $\bar U$, well-behaved states of field theory don't live in a factorized Hilbert space $\calh_U\otimes \calh_{\bar U}$.\footnote{For example, an attempt to write such a factorized state typically leads to problems like infinite energy density.  The obstacle results from  the type III property of the algebras associated to quantum field theories\cite{Araki}, and is associated with ``infinite entanglement" between the two regions; for reviews of some of their features see \cite{Haag,Yng,WittIII}.}  Instead, we consider separately and locally exciting degrees of freedom in the complementary regions by acting with localized operators, for example involving a field operator integrated against a function with support restricted to $U$.  Such operators commute at spacelike separations and can be used to describe information localized to either region, with the general statement being that there are commuting subalgebras of observables associated with spacelike separated regions, and that these define subsystems.  This is the basic formulation of locality in an algebraic approach to the subject\cite{Haag}.

\begin{figure}[t]
\sidecaption
% For example, with the graphicx style use
\includegraphics[scale=.80]{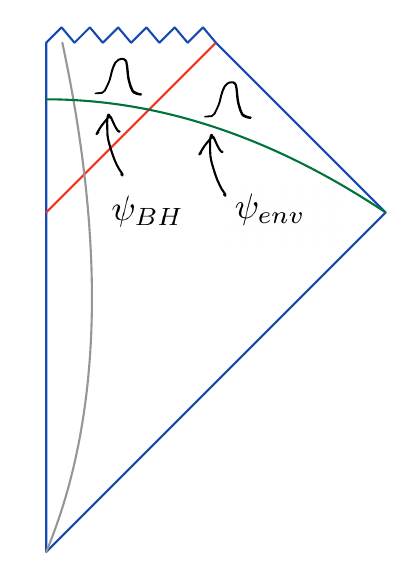}
%
% If no graphics program available, insert a blank space i.e. use
%\picplace{5cm}{2cm} % Give the correct figure height and width in cm
%
\caption{Shown is a Penrose diagram of a black hole, with one spatial slice of a family, on which we can define the state and its evolution. The state is characterized by excitations of the BH interior, and  exterior or environment, parameterized by $\psi_{BH}$ and $\psi_{env}$, respectively; in a LQFT description these can be described as excited by operators localized to the respective regions. }
\label{fig:pen}       % Give a unique label
\end{figure}

Now consider evolution of the quantum state in a black hole (BH) spacetime.  We may describe states on spacelike slices, such as shown in Fig.~\ref{fig:pen}, and write these as
\beq\label{BHstate}
|\Psi\rangle = |\psi_{\rm BH},\psi_{\rm env}\rangle\ .
\eeq
Here the labels $\psi_{\rm BH}$ and $\psi_{\rm env}$ describe the excitations localized to the black hole interior, or to the black hole environment, respectively.  For many practical purposes, one may consider modeling such a state by one in a factorized Hilbert space, which captures much of the same structure and can be written $|\Psi\rangle = |\psi_{\rm BH}\rangle \otimes |\psi_{\rm env}\rangle$.  
In the evolution described by Hawking\cite{Hawkrad}, a state that initially looks close to vacuum, {\it e.g} as measured by observers falling into the BH, evolves to one with exterior and outgoing Hawking excitations entangled with interior excitations.  A toy model for this can be written in terms of qubits,\footnote{For further discussion of such toy models, see \cite{SGqubit} and references therein.} with an initial vacuum evolving as
\beq\label{qbitevol}
|\Psi_0\rangle \rightarrow\frac{1}{2^{\frac{N}{2}}} \left( |\hat 0\rangle|0\rangle + |\hat 1\rangle|1\rangle\right)^{\otimes N}\ ,
\eeq
where hatted states denote interior excitations, and unhatted states describe the excitations of $N$ quanta of outgoing Hawking radiation.

The ``black hole theorem\cite{BHthm}" states that the following assumptions are inconsistent:

\begin{enumerate}\addtocounter{enumi}{-1}
\item{The combined system of a black hole and its environment undergoes unitary quantum evolution;}
\item{A black hole can be treated as a subsystem of the larger system including its environment;}
\item{Distinct black hole states have identical exterior evolution;}
\item{A black hole disappears at the end of its evolution.}
\end{enumerate}

This theorem can be understood as follows.  First, consider a collection of black hole states, labeled with index $i$, with the subsystem structure as described above, $|\psi_{{\rm BH}, i},\psi_{\rm env}\rangle$.  Postulate (2), identical exterior evolution, states that these evolve to later states
\beq\label{identi}
|\psi_{{\rm BH}, i},\psi_{\rm env}\rangle\rightarrow  |\Psi(t)\rangle = |\psi'_{{\rm BH}, i},\psi'_{\rm env}\rangle
\eeq
where the description of the exterior state, $\psi'_{\rm env}$ is independent of the BH state $i$.  In the LQFT description, this follows from the locality property: an internal excitation of the BH cannot influence observables outside the BH.  If the BH subsequently disappears, that then corresponds to evolution
\beq\label{disapp}
|\psi'_{{\rm BH}, i},\psi'_{\rm env}\rangle \rightarrow |\psi''_{env}\rangle\ .
\eeq
But, this is not unitary, contradicting postulate (0) -- it is not even one-to-one, and ``forgets" the information contained in the label $i$.  One could for example diagnose this by entangling the initial states $|\psi_{{\rm BH}, i},\psi_{\rm env}\rangle$ with a set of auxiliary states $|\tilde i\rangle$; the information in the BH state can be described by the von Neumann (entanglement) entropy $S_{vN}=-{\rm Tr}(\rho \ln\rho)$ of the density matrix gotten by tracing out BH states.  This entropy is preserved by \eqref{identi} but would jump to zero under evolution \eqref{disapp}; the ``missing" entanglement entropy parameterizes the magnitude of the violation of unitarity.  If one considers evolution from a black hole of initial mass $M$, the corresponding information loss is comparable to the Bekenstein-Hawking entropy,
\beq\label{bhent}
S_{BH} = \frac{A}{4G} \sim \left(\frac{M}{M_p}\right)^2\ ,
\eeq
where $G$ is Newton's constant and $M_p$ is the Planck mass, and we work in units with $\hslash=1$.

This is a crisis because violation of any of the postulates (0-4) appears to lead to a contradiction with basic physical principles.  To see this, we consider possible ways to escape this contradiction.

\subsection{Proposed alternatives}

A first proposal, originally made by Hawking\cite{Hawkunc}, is to abandon the unitary postulate of quantum mechanics.  However, it was later realized that this is extremely problematic.  In essence, information transfer and energy transfer are closely associated.  So, information loss, such as Hawking proposed, will be associated with energy nonconservation.  An argument for this was given by Banks, Peskin, and Susskind in \cite{BPS}.  How bad is the energy nonconservation?  If other principles of quantum mechanics remain intact, and locality is intact enough to allow basic effective field theory reasoning, then BH formation, evaporation, and corresponding information loss can take place as part of  virtual processes.  The natural cutoff scale for these would be of order the Planck mass, $M_p$, giving an estimated temperature for our ``vacuum" that is $T\sim M_p$ -- a preposterous departure from observation.  While it has been argued \cite{UnWa} that information loss can be disassociated from such energy nonconservation, models that do so\cite{Unmod} have similar features to models of baby universe emission.  There, it was discovered\cite{Cole,GiSt2,GiTu} that such information loss is not ``true" information loss, in that it does not persist with repeated experiments, here of black hole formation and evaporation; it is more like an initial indeterminacy in the coupling constants, that can be resolved with subsequent experiments, after which there is no more information loss.  A basic message appears to be that the principles of quantum mechanics are not so easily modified.

An obvious alternative is to drop postulate (3), disappearance, as described by \eqref{disapp}.  For example, perhaps the state information parametrized by $i$ transfers into the decay products of the BH, once its size is near-planckian and a semiclassical/perturbative description is expected to fail.\footnote{For example, this scenario is now advocated in much of the loop quantum gravity community\cite{AshBH}.}  However, this is also severely constrained by basic principles.  Since the information to be accounted for is roughly of size $S_{BH}$, encoding that in outgoing radiation of ordinary quanta would require of order $N\sim S_{BH}$ outgoing quanta.  Since the available energy is that of the near-planckian BH, each quantum can then only carry energy $E\sim M_p/S_{BH}$.  Producing such a low-energy quantum takes a time $\sim 1/E$, and so emitting $N$ such quanta would take a time $T\sim M^4/M_p^5$.  In such a scenario, BHs leave extremely long-lived remnants.

Such microscopic remnants would have masses $\sim M_p$, but would have to have an unboundedly large number of internal states, since they could form from evaporation of an arbitrarily large BH.  Such a spectrum of objects -- whether long-lived or stable -- is also extremely problematic.  Again, by basic quantum principles, these could be pair produced in ordinary physical processes, perhaps with a tiny amplitude $\cala$ for a given remnant state.  However, the total probability of production, including all the remnant states, is then $\sim N|\cala|^2$, with $N$ an unboundedly large number -- all physical processes with sufficient total energy would be unstable to unbounded production of such remnants\cite{Pres,CILAR,WABHIP}.
While there have been attempts to argue for dynamics escaping the basic effective field theory reasoning behind this\cite{Banks:1992ba,Banks:1992is,planckstar}, it does not appear that such a consistent description has been found.  Simple models\cite{CBHR} support this basic understanding.  Also, in such a remnant scenario one would also expect to have unboundedly large numbers of internal states of charged black holes, and there one can analyze a semiclassical (instantonic) production process and see how the 
unbounded spectrum would contribute to unbounded production\cite{WABHIP}.  Finally, there are also related arguments against such an unbounded spectrum based on effects of virtual such remnants\cite{Susstroub}.

If we have ruled out violating postulates (0) and (3), that leaves the possibility of some violation of postulates (1) and/or (2), which in either case would represent a departure from locality as described in LQFT.\footnote{We might think of postulate (1) as describing {\it localization} of information, and postulate (2) as associated with {\it local propagation} of information.}
It is certainly true that in a perturbative approach to quantizing general relativity, the LQFT description of locality in terms of commuting subalgebras is modified.  An important question, which will be discussed further in the next section, is the nature and role of that modification, and the nature and role of additional quantum gravitational effects beyond this.

Before further discussing locality, note that there is a significant number of other proposed scenarios for a resolution of the crisis.  Many of these now  involve some modification of conventional physics on horizon scales; one example is \cite{tHoo,tHooft:2024auh}.   A large number fall within the general category of ``massive remnant" scenarios\cite{BHMR}, where at some stage in its evolution a BH is replaced by an object that is not approximately described as having a vacuum-like horizon (or, trapped surface), but instead has some completely different structure replacing that region; a conventional analog of this is a neutron star, where the star's surface and interior replaces the BH interior.  One could consider either classical or quantum modifications to general relativity as a possible origin for such a picture.  Examples of such a general scenario are gravastars\cite{gravastar}, fuzzballs\cite{fuzzball}, earlier versions of Planck stars\cite{planckearly}, frozen stars\cite{frozen}, and presumably firewalls\cite{AMPS}, if a physical realization of the latter scenario could be found.

The difficulties with such scenarios are that they both typically also involve some nonlocal physics, {\it and} they require considerable extra physical structure that potentially creates other problems and is challenging to realize in a physically consistent scenario.  How nonlocality is invoked depends on when the massive remnant forms.  If a massive remnant is to form in a collapse before the would be horizon forms, this appears to require nonlocal physics because the local description of the collapsing matter is in a very ordinary regime, for a big black hole.  Specifically, with collapse of a gas of total mass $M$, the horizon forms when gas density reaches $\rho\sim M_p^6/M^2$, so {\it e.g.} less than the density of air in our atmosphere for a large enough black hole.  It would appear to require a nonlocal effect on the scale of the horizon to replace such a collapsing gas by an exotic object.  Alternately, such a massive remnant might form after the trapped surface of a BH had formed.  But at that point, there is a large amount of information from {\it e.g.} the collapsing matter that is inside the BH, and that information would have to be transferred at spacelike separation to the exterior as part of formation of the exotic object.  The combined difficulties of nonlocality and extra structure motivate seeking a more minimal scenario.

Finally, a very popular view is that the problems of quantum gravity, and of quantum black holes in particular, are resolved through an AdS/CFT map between gravitational physics in anti de Sitter space and conformal field theory in one fewer dimension\cite{AdSCFT}, or analogs to it with other asymptotic structure.  The problem here is understanding the detailed nature of the ``holographic map" between the CFT and the bulk quantum gravitational physics, which should in turn match onto an approximate bulk description in terms of 
 LQFT evolution and semiclassical GR.  There has been significant discussion about the origin and nature of such a map.  The most promising proposal seems to be that it is explained via essential properties of gravitational physics\cite{Maro1,Maro2}, as opposed to be due to, {\it e.g} other effects in string theory.  However, closer examination seems to show that the holographic map between ``bulk" and ``boundary" relies on solving the gravitational constraints, or their full quantum generalization, and in gravity this appears inseparable from having a description of the bulk quantum-gravitational evolution\cite{HoloUn}.  This suggests that in order to determine the holographic map, one needs to first understand the quantum-gravitational evolution in the bulk, whereas an original hope was that the boundary evolution plus holographic map determined the bulk evolution.  
This resulting circularity so far appears to preclude using the boundary theory to describe bulk evolution, and provide a resolution to the crisis that includes BH evolution that completes a bulk description.  An open question is how to provide such a precise holographic map that would give a description of evolving black holes in terms of unitary boundary evolution.

\section{Evolution and localization in local QFT and perturbative quantum general relativity}
\label{LQFTQGR}

This section turns to further description of LQFT evolution in BH spacetimes, and some of the effects of perturbative gravity.  This will add detail to the preceding discussion, begin to address questions of locality in quantum gravity, and set the stage for parametrization of possible quantum gravity effects going beyond such a description. 

\subsection{LQFT Evolution}
\label{LQFT}

\begin{figure}[t]
\sidecaption
% For example, with the graphicx style use
\includegraphics[scale=.7]{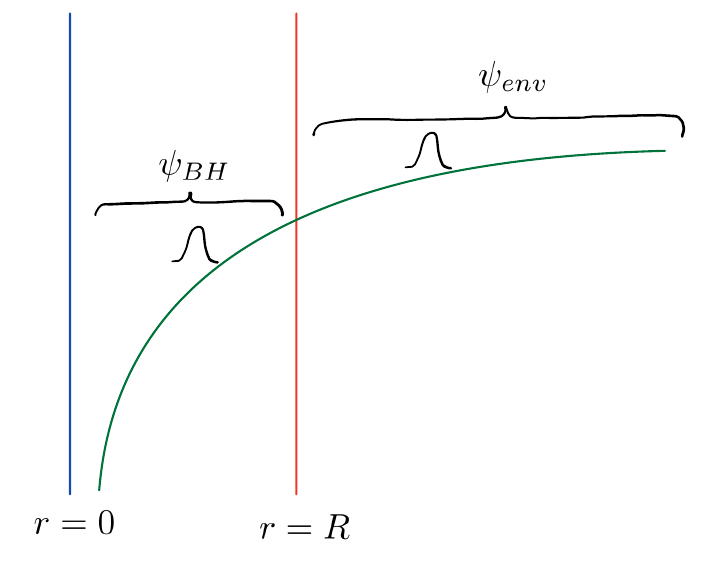}
%
% If no graphics program available, insert a blank space i.e. use
%\picplace{5cm}{2cm} % Give the correct figure height and width in cm
%
\caption{Shown is an Eddington-Finkelstein diagram for a BH, corresponding to the Penrose diagram of Fig.~1.  In such a diagram, evolution can be described by simply translating a spatial slice like that shown vertically up or down.}
\label{fig:EF}       % Give a unique label
\end{figure}

Many of the essential issues can be described in terms of evolution of a spinless scalar field on a non-spinning black hole; spin of either kind adds additional structure, but does not appear to change the basic nature of the problem.  We therefore consider  a scalar action
\beq\label{mact}
S_m=-\hf\int d^4x \sqrt{|g|}\left[(\nabla\phi)^2 + m^2 \phi^2\right]
\eeq
with a background Schwarzschild metric, initially ignoring gravitational background and fluctuations. It is useful to write the metric in ingoing Eddington-Finkelstein coordinates, $(x^+,r)$, since those are regular across the horizon, as
\beq\label{Schw}
ds^2=  -f(r)dx^{+2} +2dx^+ dr + r^2 d\Omega^2\ ,
\eeq
with $f(r)=1-R/r$ and  $R=2GM$ the Schwarzschild radius.
This geometry is pictured in Fig.~\ref{fig:pen} or equivalently by the Eddington-Finkelstein diagram of Fig.~\ref{fig:EF}.  Hawking's derivation\cite{Hawkrad} and many subsequent versions of it are based on comparing modes at future infinity to modes near the horizon or in the past, and deriving the resulting Bogoliubov transformations.  Alternately, if one wants to describe an evolving quantum state like \eqref{BHstate}, one can introduce a time slicing and a Schr\"odinger picture description of the evolution; this also has the virtue of more readily extending to interacting theories.  The choice of slicing is part of the choice of coordinates; for Schwarzschild spacetime one may choose a slice like those shown in Fig.~\ref{fig:EF}, and then define the family of slices by translating it forward or backward using the Killing vector that translates in Schwarzschild time, with the asymptotic Schwarzschild time used to parametrize the slices.  

A general slicing and metric can be described in the ADM form\cite{ADM},
\beq\label{ADMmet}
ds^2 =-N^2 dt^2 + q_{ij}(dx^i + N^i dt)(dx^j + N^j dt)\ ,
\eeq
with lapse $N$, shift $N^i=q^{ij}N_j$, and spatial metric $q_{ij}$.  
If  in the Schwarzschild metric \eqref{Schw} we choose a set of slices parametrized by a ``slicing function" $S(r)$ as
\beq\label{slicedef}
x^+ = t + S(r)\ ,
\eeq
then these ADM variables become\cite{NVU}
\beq\label{ADMvar}
N=\frac{1}{\sqrt{S'(2-fS')}}\ ,\ N_r=1-fS'\ ,\ q_{rr}=S'(2-fS')\ .
\eeq

The evolving state can be described by introducing the canonical variables and hamiltonian.  The canonical momentum is given in terms of the normal 
\beq
n^\mu =\frac{1}{N}(1,-N^i)
\eeq
to the slices as $\Pi =\sqrt q \partial_n\phi$, and the canonical commutators are
\beq
[\Pi(t,x^i), \phi(t,x^{i'})] = -i\delta^3(x-x')\ .
\eeq
Comparing the hamiltonian form of the action
\beq
S_m=\int d^4x \left(\Pi\dot \phi - \calh\right)
\eeq
 to the matter action \eqref{mact} gives the hamiltonian
 \beq\label{Hslice}
 H=\int d^3x  \calh = \int d^3x\left[ \frac{N}{2}\left(\frac{\Pi^2}{\sqrt q} + \sqrt q q^{ij} \partial_i\phi \partial_j \phi\right) + N^i \Pi \partial_i\phi\right]\ 
\eeq
and the resulting unitary evolution of the state on the slices,
\beq\label{stateevol}
|\Psi(t)\rangle = e^{-iH(t-t_0)}|\Psi(t_0)\rangle\ .
\eeq 

The structure \eqref{BHstate} of the state is found by beginning with an appropriate fiducial ``vacuum" state on the slice, and considering excitations of it involving combinations of the operators $\phi, \Pi$ localized either to the BH interior, or to the exterior.\footnote{A more careful statement involves using unitary operators that are localized in the respective regions.}  For example, different such operators localized inside the BH create BH states with different labels $i$, as described in \eqref{identi}.  Because of the locality property of the QFT, described above in terms of commuting subalgebras, different such operators commute with observables exterior to the BH on subsequent slices; the corresponding excitations propagate within the lightcone, and so these excitations don't affect the external part of the state.

In addition, the evolution \eqref{stateevol} describes creation of correlated pairs of excitations inside and outside the BH, like the toy model evolution \eqref{qbitevol}.  This can be seen by studying the explicit evolution \eqref{stateevol} written in terms of creation/annihilation operators, using the explicit form of the ADM variables \eqref{ADMvar}, as was done in two dimensions in 
\cite{SEHS,SE2d} and for higher-dimensional BHs in \cite{GiPe1}.  One finds that the hamiltonian \eqref{Hslice} has a term bilinear in creation operators, which creates the correlated pairs.  
This leads to a refinement of the evolution law\eqref{identi}, 
\beq\label{evolref}
|\psi_{{\rm BH}, i},\psi_{\rm env}\rangle\rightarrow  |\Psi(t)\rangle =\sum_\alpha |\psi'_{{\rm BH}, i},\psi'_{{\rm BH}, \alpha}; \psi'_{\rm env,\alpha}, \psi'_{\rm env}\rangle
\eeq
in which the exterior is still independent of $i$, but has built up additional entanglement with other interior degrees of freedom, labelled by $\alpha$, through subsequent evolution.
The specific form of the hamiltonian depends on the choice of  slicing and coordinates on the slices, which correspond to a choice of gauge in the gravitational context.

This basic structure appears to extend to spinning BHs, with additional complications of detail.  While there are additional subtleties in describing  evolution for time-dependent geometries and/or slicings\cite{Helf,ToVa1,ToVa2,AgAs,GiPe3}, it nonetheless appears that such unitary and local evolution can still be defined.

\subsection{Perturbative quantized general relativity (QGR)}

A critical question is what gravitational corrections modify the preceding story, and how they do so.  

Gravitational backreaction is an essential part of the discussion, since it describes the shrinkage of the BH.  A first approach to this is by studying the average backreaction as a correction to the semiclassical metric.  This has in particular been done in two-dimensional models, and specifically in dilaton gravity in \cite{CGHS} and many subsequent references.  These provide models which are also expected to capture basic features of higher-dimensional evolution, since for large BHs this evolution is dominantly spherically symmetric, and s-wave dynamics can be described in such a two-dimensional theory.  Specifically one does also explicitly find buildup of a large entanglement/missing information in such models describing a BH shrinking towards zero mass, as was shown in \cite{GiNe}, and so at this level backreaction appears not to greatly alter the discussion of the preceding section.

The situation is more subtle at the quantum level.  To see this, we can investigate the leading quantum corrections to the classical or semiclassical evolution perturbatively in Newton's constant $G$.  The gauge symmetry of gravity now plays an important role, and in particular the basic field operators $\phi(x)$ are no longer observables, since they are not gauge invariant.  When one replaces them by gauge-invariant observables {\it e.g.} by including {\it gravitational dressing}, these are no longer local, and the preceding formulation of locality and subsystems in terms of commuting sublagebras of observables apparently must be modified.

There isn't space in this review for a complete discussion of gravitational observables, dressing, and subsystems, and significant open questions remain.  But, it is important to take some time to summarize some of the basic features of such a discussion, with further details provided in \cite{DoGi1,DoGi2,DoGi3,DoGi4,SGsplit,GiKi,GiPe2}, since this drives at the very important question of understanding localization in quantum gravity, and connecting with the discussion of Sec.~\ref{bhthm}.

\subsubsection{Perturbative gravitational dressing}

One can formally describe the evolving state by extending the hamiltonian framework to include gravity.  A starting point is the lagrangian
\beq\label{gravact}
S=\frac{1}{16\pi G}\int d^4x \sqrt{|g|} \calr +S_\partial +S_m\ ,
\eeq 
where $\calr$ is the Ricci scalar, and $S_\partial$ is a boundary term.\footnote{For further details of the following treatment, see \cite{GiPe2}.}
Canonical variables are defined using the ADM decomposition \eqref{ADMmet}.  Rewriting the action \eqref{gravact} in these variables, the canonical momentum becomes
\beq
P^{ij} = \frac{\delta S}{\delta \dot q_{ij}} = -\frac{1}{16\pi G} \sqrt q \left(K^{ij}-q^{ij} K\right)
\eeq
where dot denotes $\partial/\partial t$ and $K_{ij}$ is the extrinsic curvature of the slices,
\beq
K_{ij}=\frac{1}{2N} \left(-\dot q_{ij} + D_iN_j + D_j N_i\right)\ ,
\eeq
with $D_i$ the covariant derivative defined from $q_{ij}$.  
This is taken to satisfy commutation relations
\beq\label{gccr}
[P^{ij}(t,x) , q_{kl}(t,x')]= -\frac{i}{2}\left(\delta_k^i \delta_l^j + \delta_l^i \delta _k^j\right)\delta^3(x-x')\  .
\eeq
The momenta conjugate to $N, N^i$ vanish, in association with the presence of gauge constraints.  

Comparing the hamiltonian form of the action
\beq
S= \int d^4x \left( P^{ij} \dot q_{ij} + \Pi \dot \phi - \calh\right)
\eeq
to the action \eqref{gravact} yields the hamiltonian in two different equivalent forms,
\beq\label{fullham}
H=\int d^3 x\left(N\calc_n + N^i \calc_i \right) + H_\partial = \int d^3x \left(\calh_g + \calh_m\right)\ .
\eeq
In the first expression, we define
\beq
\calc_\mu = \sqrt q\left(-\frac{G_{\mu\nu}}{8\pi G} + T_{\mu\nu} \right)n^\nu\ ,
\eeq
and have used the index convention $\calc_n=n^\mu\calc_\mu$,
with Einstein tensor $G_{\mu\nu}$ and stress tensor $T_{\mu\nu}$.  Vanishing of the $\calc_\mu$ expresses the constraints associated with the gauge symmetry; correspondingly, $N,N^i$ appear as Lagrange multipliers in the expression.  $H_\partial$ is a boundary term at spatial infinity; {\it e.g.} in the case of Minkowski asymptotics it reduces to the ADM boundary hamiltonian.  Since the constraints have second derivatives of the metric variables, the latter term is needed so that after integration by parts one has a hamiltonian of the standard form with just quadratic terms in momenta and the spatial derivatives of the fields, together with interaction terms.  The second expression with $\calh=\calh_g+\calh_m$ summarizes that form, with $\calh_m$ the matter hamiltonian, and $\calh_g$ an effective gravitational  hamiltonian.

When the constraints $\calc_\mu$ vanish, the hamiltonian thus reduces to a pure boundary term.  Since the $\calc_\mu$
generate the gauge symmetries (diffeomorphisms) on shell,\footnote{For an explicit discussion, see \cite{GiPe2}.} their vanishing is an expression of gauge invariance.
Formally the evolution of the state  $|\Psi\rangle$  is then determined by the Schr\"odinger equation
\beq\label{Scheq}
i\frac{\partial}{\partial t} |\Psi\rangle = H |\Psi\rangle\ 
\eeq
and is nontrivial if we for example fix the time parameter $t$ at infinity, like we have done in \eqref{slicedef}, \eqref{ADMvar}.
If the second expression in \eqref{fullham} is used, \eqref{Scheq} takes a standard and unsurprising form, with both the matter and gravity hamiltonians contributing to evolution. 
But for a state annihilated by the constraints, this evolution could alternately be expressed in terms of the action of the boundary hamiltonian $H_\partial$.\footnote{The actual story is a little more complicated since typically one only wants ``half" of the constraints to annihilate the state; then the remainder are expected to generate contributions that decouple.  Some further discussion of this appears in \cite{DoGi4,GiPe2}.}

For operators, the condition of gauge invariance is expressed as\footnote{Again, modulo the equations of motion.} 
\beq\label{giop}
[\calc_\mu,\calo]=0\ 
\eeq
Analogous to the discussion in \ref{LQFT}, we can then consider creating different states by acting with different such operators $\calo_i$ on some fiducial physical state $|\Psi_0\rangle$.   However, in gravity gauge invariant operators cannot be local\cite{Torre}.  Correspondingly, solving \eqref{giop} will yield {\it nonlocal} operators, and these will generically not satisfy the local algebra of LQFT; locality in gravity must apparently be defined some other way\cite{SGalg,QFG,QGQF}

The preceding discussion is somewhat formal, and there are significant challenges in solving the constraints and determining the evolution in the full nonlinear theory.  However, one can begin to understand the corresponding structures by working perturbatively in $G$, and in particular considering the leading perturbative contributions.  These are expected to dominate the behavior for weak gravitational fields, and for example should therefore capture the leading long-distance behavior, since the field falls off at long distances.
 This leading behavior already nontrivially illustrates some of this novel gravitational structure, as well as its modification to the discussion of the preceding sections.

For example, one may begin with an operator $\calo$ that is a purely matter operator, and add gravitational contributions -- that is, ``gravitationally dress" it -- to make an operator satisfying \eqref{giop}.  Such gravitationally dressed operators have been constructed in \cite{DoGi1,DoGi2,DoGi3,QGQF,DoGi4,GiKi},\footnote{Earlier related work includes \cite{Heemskerk} and \cite{KaLi}. The first derived related nontrivial commutators as arising from the
constraints, but didn't give the dressed operators, and the second focussed on deriving commuting bulk operators.} and recently in nontrivial backgrounds such as BHs in \cite{GiPe2}.  To construct these we perturb about a background metric $g_{\mu\nu}$,
\beq
\tilde g_{\mu\nu}=g_{\mu\nu} +\kappa h_{\mu\nu}\ ,
\eeq
with $\kappa^2=32\pi G$, and define momenta perturbations,   over background momenta $P^{ij}$,
\beq
\tilde P^{ij}=P^{ij}+ p^{ij}/\kappa
\eeq
so that $h_{ij}, p^{ij}$ also satisfy the commutators \eqref{gccr}.
To leading nontrivial order in $\kappa$, we can take a matter operator $\calo(t,x)$ or a more general nonlocal operator $\calo(t)$, and then its gravitationally dressed version is
\beq\label{dressop}
\hat \calo_t= e^{i\int d^3x \sqrt q V^\mu(t,x) T_{n\mu}} \calo(t) e^{-i\int d^3x \sqrt q V^\mu(t,x) T_{n\mu}}\ ,
\eeq
where the gravitational dressing $V_\mu$ is at this order linear in $h_{ij}$ and $p^{ij}$.  Specific such expressions are given in the preceding references, and take the general form\cite{GiPe2}
\beq\label{dressvn}
V^n(x)=\frac{\kappa}{2} \int d^3x'\left[\breve h_{ij}(x,x') p^{ij}(x') - \breve p^{ij}(x',x)h_{ij}(x')\right]\ ,
\eeq
\beq\label{dressvi}
V^i(x)=\kappa\int d^3 x' \left[ G^{ijk}(x',x) h_{jk}(x') + H^{i}_{jk}(x',x) p^{jk}(x')\right]
\eeq
where $\breve h_{ij}$, $p^{ij}$, $G^{ijk}$, $H^{i}_{jk}$ are c-number Green functions satisfying a corresponding set of equations.
For a given matter configuration, there are different dressings corresponding to different solutions of the Green function equations; a simple example is to dress a localized mass with either a Coulomb-like or a line-like gravitational dressing\cite{DoGi1}.

Since, with the example of Minkowski or AdS asymptotics, the support of the dressing generically extends to infinity\cite{DoGi2,GiKi}, the operators \eqref{dressop} based at different spacelike separated points generically don't commute\cite{SGalg,DoGi1}, concretely illustrating the preceding general comments about modification of the LQFT algebra.

\subsubsection{Gravitational splitting and localization of information}

While this raises the question of how in general to define locality and subsystems of gravitational systems, there is a perturbative construction that is close to recovering the discussion of the preceding sections.  
This nonetheless leaves important open questions about localization of information in the full theory of quantum gravity.  
The question is most easily illustrated by perturbing about a flat metric, but the more general description of dressing outlined above also serves as the starting point for extending the discussion to nontrivial background metrics such as BHs.  
This construction was called a ``gravitational splitting" in \cite{DoGi3,DoGi4,SGsplit}.

The starting point is the essential question in quantum gravity of whether information can, to a good approximation, be described as localized to a region -- for example interior to a BH -- like it is in LQFT, as described above.  Or, do gravitational effects  lead to important modifications to this, or, serve to transfer the information out of the region?

We will investigate this question in perturbative QGR.\footnote{For further discussion see \cite{SGasymp}.}  A useful toy model to do so arises from considering two degenerate scalar fields, $\phi_a=(\phi_1,\phi_2)$, governed by a lagrangian which is two equal-mass copies of \eqref{mact}, which has a 
global symmetry $\phi_a\rightarrow O_{ab}\phi_b$ relating the fields.  Of course there is significant doubt about the existence of such global symmetries in a full quantum theory of gravity, but it so far appears consistent to consider the perturbative coupling of such a theory to gravity.  Such a toy model then furnishes an example of what are plausibly key features of information localization in a more complete description of the states in a more complete theory.

We begin with a fiducial state $|\psi_0\rangle$ -- which in the Minkowski case can just be the Minkowski vacuum $|0\rangle$ -- and consider excitations of it that describe information localized in a region $U$, such as a small open set.  In QFT we take these to be created by unitary operators $U_I$ acting on $|\psi_0\rangle$, $|\psi_I\rangle = U_I|\psi_0\rangle$.  
For concreteness consider examples such as 
\beq\label{expsource}
U_I= e^{-i \int d^3x J^I_a(x) \phi_a(x)}\ ,
\eeq
where the functions $J^I_a$ have compact support restricted to $U$; or one could consider more general localized unitary operators.  These states have the property that for any local operator $\calo(x')$, with $x'$ spacelike separated from $U$,
\beq
\langle\psi_I|{\cal O}(x')|\psi_I\rangle=
\langle\psi_0| U_I^\dagger \calo(x') U_I|\psi_0\rangle = \langle\psi_0|  \calo(x')|\psi_0\rangle \quad :
\eeq
measurements of local observables spacelike to $U$ are insensitive to the state created in $U$.  A concrete example is given by the functions $J^a_b(x) = \delta^a_b J(x)$, with a common $J(x)$ of compact support -- measurements outside $U$ do not distinguish whether one acted using $J^1$ or $J^2$, that is which ``color" of particles are present in the state.

This discussion must be modified in quantum gravity, since the operators $U_I$ are not gauge invariant.  We can make them gauge invariant, at least  to leading order in $\kappa$, using a dressing $V_\mu(x)$ as described above,
\beq
\label{uihatdef}
\hat U_I = e^{i\int d^3x  V^\mu(x) T_{n\mu}}U_I e^{-i\int d^3x  V^\mu(x) T_{n\mu}}\ .
\eeq
As in \eqref{dressvn}, \eqref{dressvi}, the dressing extends outside $U$, potentially affecting measurements there.  

\begin{figure}[]
\sidecaption
% For example, with the graphicx style use
\includegraphics[scale=.90]{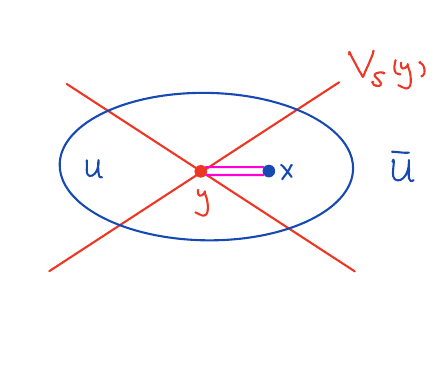}
%
% If no graphics program available, insert a blank space i.e. use
%\picplace{5cm}{2cm} % Give the correct figure height and width in cm
%
\caption{Illustration of the standard dressing construction.  A standard dressing $V_S(y)$ is chosen that only depends on the neighborhood $U$ and {\it e.g.} a given point within it.  One may choose different forms for this dressing, {\it e.g.} line-like or Coulomb-like.  Dressing of operators within $U$ is then constructed by dressing from a general point $x\in U$ to $y$, as illustrated; an example construction is via a line-like dressing.  The result is the dressing expression \eqref{Sdress}. }
\label{fig:stan}       % Give a unique label
\end{figure}

However, as noted above there are different possible dressings, and we may use this freedom to find gravitational states minimizing the information ``seen" from outside $U$.  This can be done by constructing\cite{DoGi3,DoGi4,SGsplit} a ``standard dressing" associated to the neighborhood $U$ itself.  Given such a dressing $V^\mu_S$, we may write
\beq\label{Sdress}
V^\mu(x) = V^\mu(x,y) + V^\mu_S(y) +\hf(x-y)_\nu\left[\partial^\nu V_S^\mu(y) - \partial^\mu V_S^\nu(y)\right]\ .
\eeq
Here $y$ is a specified point of $U$, and the dressing $V^\mu(x,y)$ has the property that it has no support outside a bounded set containing both $x$ and $U$ -- for $x\in U$, it may be taken to vanish outside $U$.  Colloquially, we say that we dress from the point $x$ to $y$ -- which for example may be accomplished via a line dressing like in \cite{DoGi1,QGQF} -- and then the dressing of the point $y$ extends to infinity; see Fig.~\ref{fig:stan}  .

Given such a dressing, consider measuring an operator $\calo'$ in the now quantum-gravitational state\footnote{The state $|\psi_0\rangle$ may also need dressing, which we do not make explicit.  Also, more generally, $\calo'$ may be dressed, {\it e.g.} to infinity.}   $|\hat \psi_I\rangle=\hat U_I |\psi_0\rangle$.  We find
\beq
\langle\hat\psi_I|\calo' |\hat\psi_I\rangle=
\langle\psi_0| \hat U_I^\dagger \calo' \hat U_I|\psi_0\rangle = \langle\psi_0|  \calo'|\psi_0\rangle + \langle\psi_0| \hat U_I^\dagger [\calo',\hat U_I]|\psi_0\rangle\ ,
\eeq
where, for $\calo'$ localized in the spacelike complement to $U$, the leading order in $\kappa$ contribution to the difference term may be evaluated  by a straightforward commutator calculation, using \eqref{uihatdef}, to yield
\beq\label{oshift}
\delta \langle\hat\psi_I|\calo' |\hat\psi_I\rangle=\langle\psi_0| \hat U_I^\dagger [\calo',\hat U_I]|\psi_0\rangle \simeq i \int d^3x \langle\psi_0| U_I^\dagger \left[\calo',V^\mu(x)\right] \left[T_{0\mu}(x),U_I\right] |\psi_0\rangle)\ .
\eeq

For example, consider the case where $\calo'$ is purely constructed from the gravitational perturbation.  Then the latter expression factorizes,
\beq
 \langle\psi_0| U_I^\dagger \left[\calo',V^\mu(x)\right] \left[T_{0\mu}(x),U_I\right] |\psi_0\rangle =  \langle0| \left[\calo',V^\mu(x)\right]|0\rangle \langle\psi_0| U_I^\dagger \left[T_{0\mu}(x),U_I\right] |\psi_0\rangle \ ;
\eeq
the second factor has support only for $x\in U$, but then the first factor receives no contribution from $V^\mu(x,y)$ in \eqref{Sdress} and so only has $x$-dependence from the standard dressing parts of \eqref{Sdress}.  As a result, the integral over $x$ just yields Poincar\'e generators, and
\bea
\delta \langle\hat\psi_I|\calo' |\hat\psi_I\rangle &\simeq& - i \langle\psi_0| U_I^\dagger \left[P_\mu, U_I\right]|\psi_0\rangle\ \langle0| \left[\calo',V_S^\mu(y)\right]|0\rangle\cr
 &&-\frac{i}{2} \langle\psi_0| U_I^\dagger \left[M_{\mu\nu} ,U_I\right]|\psi_0\rangle\ \langle0| \left[\calo',\partial^\mu V_S^\nu(y)\right]|0\rangle\ .
\eea

This expression depends on the state created by $U_I$ only through the wavefunction for its total Poincar\'e charges.  For example in our model with global symmetry, this tells us that measurement of gravitational operators outside $U$ do not distinguish the color of the state created by $U_I$; more generally, states with equal Poincar\'e moments will yield the same measurements of $\calo'$.\footnote{In our model we achieve such an exact degeneracy with the global symmetry. While there are indications that a complete theory of gravity may not have such global symmetries, there may be degeneracies or very near degeneracies in Poincar\'e charges which are in-practice unresolvable via measurements outside $U$. } This basic construction was called a ``gravitational splitting" in \cite{DoGi3,DoGi4,SGsplit}.  One can understand this as a perturbative quantum generalization of certain classical constructions, where the classical gravitational field outside a given region may be taken to be of a specified form, {\it e.g.} Kerr or a cone-like distribution\cite{CoSc,CaSc}, depending only on the momentum and angular momentum within the region.

An interesting modification of this story arises if $\calo'$ is a mixed matter/gravitational operator; consider the simple example of a product
\beq
\calo' = \calo'_h \calo'_\phi\ .
\eeq
In this case, \eqref{oshift} becomes 
\beq
\delta \langle\hat\psi_I|\calo' |\hat\psi_I\rangle\simeq i \int d^3x\langle0| \left[ \calo'_h,V^\mu(x)\right]|0\rangle  \langle\psi_0| U_I^\dagger \calo'_\phi \left[T_{0\mu}(x),U_I\right]  |\psi_0\rangle\ .
\eeq
Again the second factor is only nonvanishing for $x\in U$, and so only the standard dressing terms in the first commutator contribute.  The integral over $x$ then again yields Poincar\'e generators, with the result
\bea
\delta \langle\hat\psi_I|\calo' |\hat\psi_I\rangle&=& - i \langle\psi_0| U_I^\dagger\calo'_\phi \left[P_\mu, U_I\right] |\psi_0\rangle\ \langle0| \left[\calo',V_S^\mu(y)\right]|0\rangle\cr
 &&- \frac{i}{2} \langle\psi_0| U_I^\dagger\calo'_\phi \left[M_{\mu\nu} ,U_I\right]|\psi_0\rangle\ \langle0| \left[\calo',\partial^\mu V_S^\nu(y)\right]|0\rangle\ .
\eea
Now, the measurement {\it can} depend on details of the state.  For example in the toy model, if one takes $\calo'_\phi = \phi_1(x')$, the first correlator depends on which color particles $J^I_a$ creates\cite{SGasymp}.  However, the non-vanishing result arises from a correlator $\langle\phi_a(x)\phi_b(x')\rangle$, and so is exponentially suppressed $\sim \exp\{-m|x-x'|\}$ for $x'$ spacelike to $U$.  

The presence of such nonvanishing correlators is certainly interesting, but due to their smallness, what is not clear is their relevance for significantly altering the story of localization and transfer of information.\footnote{This is despite arguments to the contrary\cite{CGPR}.}  These small corrections have only been calculated to leading order in $\kappa$, and so there are additional corrections to account for at higher orders.  Moreover, one can ask how the gravitational dressing modifies the evolution of the state in a perturbative description.  While in the perturbative description one does find that the state \eqref{stateevol} must be dressed, there is no obvious indication that this alters the internal degrees of freedom of the BH, or the pattern of entanglement between them and the exterior degrees of freedom, as described {\it e.g.} in \eqref{evolref}.  So, for practical purposes it so far seems that evolution of a BH as a subsystem like in the non-gravitational LQFT description is a good approximate description.

Interestingly, at the nonperturbative level related arguments formally extend to the leading proposal for the explanation of the property of ``holography," as associated with solving the gravitational constraints.  Such an explanation was first given by Marolf\cite{Maro1,Maro2}, who argued that an operator, say $\phi(x)$, in the middle of AdS could be related to an operator in its future lightcone on the boundary of AdS, but then since the hamiltonian can be thought of as a boundary term, the latter operator can be translated back in time in the boundary algebra to give a boundary operator at equal time to $\phi(x)$ -- constructing a ``holographic map"  between bulk and boundary operators.  Crucially, this depends on solving the gravitational constraints in AdS\cite{HoloUn}, which amounts to constructing the dressing, since that is needed to express the hamiltonian as a pure boundary term, as seen in \eqref{fullham}.

Indeed, a related but simpler formal description of a possible holographic map was given in \cite{DoGi3}.  Suppose, now, that $|\hat\psi_I\rangle$ are  fully-nonperturbative dressed versions of the states $\hat U_I|\psi_0\rangle$ that were described above.  From a momentum analog of the formula \eqref{fullham}, one expects that for states annihilated by the constraints, the translation generators are given by their boundary, ADM, versions.\footnote{There are additional subtleties with this which we leave for other discussion.}  This means that 
\beq
e^{i P_i^{ADM} c^i} |\hat\psi_I\rangle
\eeq
should translate the state by the amount $c^i$; {\it e.g.} in the case of $U_I$ given by \eqref{expsource} this is equivalent to translating the support of $J(x)$.  For large $c^i$ we can translate to ``near to the boundary," and so for example the expression
\beq
\langle \hat\psi_I | \phi_i(x') e^{i P_i^{ADM}c^i }|\hat\psi_I\rangle\ ,
\eeq
for $x'$ near to the boundary, is sensitive to details of the state.  By taking a limit of infinite $c^i$ and $x'$, this is the starting point for an argument that bulk states can be detected -- and created -- by boundary operators.

However, this formal argument appears to require a full solution of the constraints, presumably with nonperturbative accuracy, since one must implement a large translation to construct the holographic map.  Since solving the constraints is equivalent to finding the bulk evolution, this appears to indicate that in order to construct the holographic map one must first understand the bulk evolution -- and in particular have a consistent description of black hole evolution\cite{HoloUn}.  The resulting apparent circularity in the argument thus leaves its status, particularly with somehow resolving the unitarity crisis, in question.

Returning to the question of  BH evolution, the black hole theorem appears to tell us that there need to be effects that modify localization and/or transfer of information in the BH context.  The upshot of the preceding discussion is that in a leading perturbative analysis, there is so far no clear indication that the relevant effects arise.  One finds a description of the evolution that to leading order includes the buildup of entanglement of Hawking's picture, and at next order in $G$ incorporates gravitational dressing of the resulting entangled states; there is not a clear indication for elimination or significant rearrangement of the underlying entanglement.
This suggests that there must be effects outside such a perturbative description that modify the BH evolution.  

Whether such effects arise from some nonperturbative completion of the dressing construction, or from some other fundamental picture that goes beyond quantized GR, we will now shift to a pragmatic view: we will seek to give a principled ``parameterization of ignorance" of effects that can resolve the unitarity crisis.  Ultimately, one would like to derive such effects as arising from a more fundamental description of quantum geometry, but for the time being we instead take such an ``effective" approach, in the spirit of effective field theory descriptions that have been successful in many other contexts.

\section{Nonviolent unitarization: motivation and basic approach}
\label{NVU}

To briefly summarize the preceding motivations, we have outlined that there are significant difficulties altering quantum mechanics.  In order to instead save it, specifically its principle of unitarity, while avoiding other disastrous behavior, we apparently need some modification to the QFT description of locality of information, and/or apparently more radical measures.  

The latter typically involve considerable additional structure which is problematic from multiple viewpoints, and moreover black hole  observations so far appear to match classical predictions for BH behavior.  Since LQFT has served to give an excellent and well-tested description of all other non-gravitational phenomena, these together
help motivate taking a conservative approach of seeking the ``minimal" modifications to the LQFT description of BH evolution, as it was reviewed above, that are needed to restore unitary evolution.  One facet of this is that we seek a consistent description of BHs that involves a minimal modification to the equivalence principle.  These are key aspects of the general proposal of ``nonviolent unitarization\cite{SGmodels,BHQIUE,GiSh1,NVNL,NVNLFT,GiSh2,NVNLT,NVU,BHQU}" (NVU).

A starting point for discussing NVU is therefore the LQFT evolution; it appears easiest to begin with a hamiltonian description of it, as summarized above.  We then explore the question of what corrections to this picture, that are in a certain sense ``small," are needed to restore unitarity to the complete BH evolution. 
As also summarized above, the LQFT evolution produces increasing entanglement between internal states of the BH and the outgoing radiation.  If the more complete gravitational description permits, at least to the necessary approximation, a description of the BH as a subsystem that contains information and becomes entangled with the environment as in the preceding leading order description, and if we accept postulates (0), ultimate unitarity and (3), no remnants, for the reasons described above, then the black hole theorem tells us that we need a violation of postulate (2): distinct BH states should {\it not} have identical exterior evolution, or, put differently, there should be some interaction between the BH states and the BH exterior.  This must apparently be outside the realm of LQFT, but we will seek to parametrize it as a departure from LQFT.

In doing so, we could begin with the leading-order corrected version of the LQFT description, where states and observables have the leading-order gravitational dressing described above.  However, since that has not been found to yield the needed modifications to the original uncorrected LQFT description, and produces extra complications, we will instead work with a description that neglects these effects, and focus on other corrections to the evolution that are needed.

Specifically, in such a description, we will take a ``principled" approach of parameterizing our ignorance, seeking the minimal modifications to the underlying LQFT evolution that are needed to restore ultimate unitarity.

\section{Parametrizing nonviolent unitarization}
\label{PNVU}

Given the preceding discussion, our effective description begins with the hamiltonian formulation, {\it e.g.} with hamiltonian \eqref{Hslice} (or, more generally, \eqref{fullham} with gravitational corrections), which we now refer to as $H_0$; we will parameterize corrections to this that can restore ultimate unitarity.  Of course this hamiltonian depends on a choice of slicing, or gauge, but in taking this  effective approach we will explore the possible interactions in such a specific gauge, and defer the question of gauge invariance to a later time when we have a more complete understanding of the fundamental underlying dynamics.  

\begin{figure}[h]
\sidecaption
% For example, with the graphicx style use
\includegraphics[scale=.70]{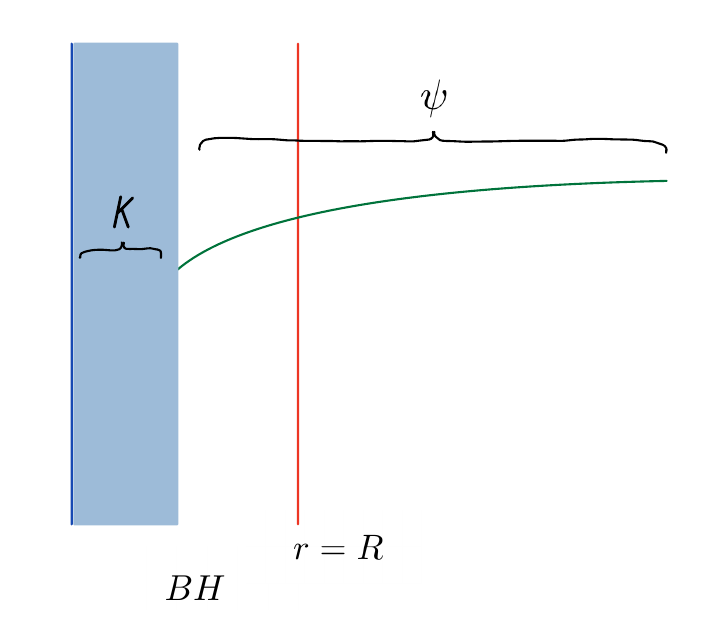}
%
% If no graphics program available, insert a blank space i.e. use
%\picplace{5cm}{2cm} % Give the correct figure height and width in cm
%
\caption{Schematic of the state as parameterized in \eqref{fullstate}.  At sufficiently large radius, including a region interior to the horizon, we expect that to a good approximation the LQFT description furnishes a good description of the degrees of freedom.  Degrees of freedom deeper in the BH are expected to receive larger corrections to their description, and will be parameterized by a state label $K$.}
\label{fig:NVUstate}       % Give a unique label
\end{figure}

We  also expect the states to be modified from their LQFT description.  However, plausibly the modification to their structure is minimal  in the vicinity of the BH horizon, and they are for practical purposes well-parametrized there by the LQFT description, in alignment with our approach of minimality.  We do anticipate, though, that there could be quite significant modification to the description of the degrees of freedom (DOF) deep inside the BH, and also that the description there depends significantly on the choice of gauge which extends the classical slice dependence.  Concretely, we could parametrize the state at a given time $t$ as (see also Fig.~\ref{fig:NVUstate})
\beq\label{fullstate}
|M; K; \psi, t\rangle\ .
\eeq
Here $M$ is the mass of the BH; in practice we may want to describe BH states in a mass range $(M+\Delta M, M-\Delta M)$, with say $\Delta M\sim 1/R$, with $R$ the horizon radius  (spin could also be included, but will be neglected for simplicity).  $\psi$ gives the description of the state of the environment, outside the horizon, as well as the near-horizon interior,\footnote{To describe localized states, one may for example work in a wavepacket basis.} say inward to some modest fraction of  $R$, 
and $K$ are labels of DOF in the deeper interior of the BH, where significant deviation from LQFT is expected.  Quite possibly the internal states are finite in number, and so for the mass range $(M+\Delta M, M-\Delta M)$, we have $K=1,\cdots,N$, where $N=\exp\{S_{bh}\}$ for a BH entropy $S_{bh}$ that may, for example, be comparable to the Bekenstein-Hawking value $S_{BH}=A/4G$.\footnote{Being more careful, one should include internal excitations from $\psi$ in the count of internal BH states.}

We then parametrize the full (effective) hamiltonian as
\beq
\label{Heffect}
H= H_0 + \Delta H\ ,
\eeq
where again $H_0$ is given by the LQFT expression \eqref{Hslice}, or in a more complete treatment, its perturbative QGR completion \eqref{fullham}.  Once again, the LQFT expression for $H_0$  respects postulate (2), identical exterior evolution, and so to violate this, and restore ultimate unitarity, $\Delta H$ must incorporate interactions between the BH internal DOF and those of its environment.  $\Delta H$ may also include significant corrections to the evolution of the internal DOF -- say as they reach near the would-be singularity.  
We will separate these two contributions as
\beq
\label{Hsep}
\Delta H= \Delta H_I + \Delta H_{bh}\ ,
\eeq
respectively.
So, $\Delta H_{bh}$ acts on the DOF labeled by $K$ in \eqref{fullstate}, and may be large, and $\Delta H_I$ couples the internal DOF to those of the BH exterior; for minimality, it should in a sense to be determined be small.

The simplest interaction that can transfer information or entanglement between two subsystems -- here BH to environment -- is a product of operators acting on the respective DOF.  A basis of such operators acting on the states labeled by $K$ is given by the $U(N)$ generators $\lambda^A$, and these can be coupled to different operators acting on the part of the state described by $\psi$.\footnote{We allow for couplings between the BH deep-interior states and the environment as well as to  the near-horizon interior, since we do not necessarily expect a sharp transition in behavior at the horizon.  However, only the pieces coupling to the environment DOF are relevant for the information transfer to the environment that violates postulate (2).}  For simplicity and minimality we take the latter to be superpositions of local operators $\calo^b(x)$, although more general nonlocal operators could be considered.  Then, the information-transferring interaction takes the form\cite{NVNLFT,GiSh2,NVNLT,NVU}
\beq\label{DeltaHI}
\Delta H_I = \sum_{Ab}\int dV_x \lambda^A \calo^b(x) G_{Ab}(x)\ ,
\eeq
where the integral is over the timeslice of the environment and near-horizon interior, and $dV_x$ is the volume element induced by the background metric.  The functions $G_{Ab}(x)$ are parameters of our effective description, behaving somewhat like ``structure functions" of the BH states, which we expect to ultimately be determined by a more complete fundamental theory but for now we regard as parametrizing our ignorance.  We will aim to find constraints on these based on other, general, considerations.

Of course when described from the perspective of the semiclassical BH geometry, \eqref{DeltaHI} is nonlocal; such a violation of the semiclassical notion of locality is again apparently {\it required}, as part of violating postulate (2).  When described from the full quantum gravity perspective, the situation is less clear, since we currently lack a complete understanding of locality in that context; perhaps, in a sense, the interactions \eqref{DeltaHI}  are part of parameterizing the departure of quantum geometry and dynamics from its semiclassical approximation.  Of course departures from locality are also {\it potentially} problematic.  For example, in Minkowski space, a nonlocal interaction can be converted into one that acts backward in time by a boost, and then such interactions can be combined to send signals into the backward lightcone of an observer.  This violates causality, and produces paradoxes such as the ``grandmother paradox."  However, in the present context, in which we work about the BH background, such apparently nonlocal interactions cannot necessarily be converted into interactions acting backward in time\cite{NLvC,BHQIUE}, since the background does not have the full Lorentz symmetry and in effect chooses a frame.  This suggests that such inconsistencies can  be avoided.

\subsection{Scales and nonviolence}\label{scaless}

This discussion connects to the question of the spatial dependence of the $G_{Ab}(x)$ -- for example if they had support to large $x$, that could parametrize arbitrarily long-range nonlocality, which {\it is} expected to produce paradoxes.  But if the new interactions are associated with the quantum structure of the BH, a natural assumption is that they extend a distance $\Delta R_a$ outside the horizon, where $\Delta R_a$ is not greatly larger than $R$.  It has sometimes been presumed that $\Delta R_a$ is microscopically small, {\it e.g.} with a scale set by the Planck length.  However, such a rapid variation of the $G_{Ab}$ would then produce very violent behavior, for example as seen by infalling observers\cite{SGTrieste,BPZ,AMPS}; the latter reference coined the name ``firewall" for the resulting object.  This, for example, represents a large violation of the equivalence principle, since an infalling observer would experience large interactions in a region where small curvatures are expected classically.  Such behavior may be avoided if $\Delta R_a$ instead scales with $R$, as
\beq\label{Radef}
\Delta R_a\sim R^p\ 
\eeq
for $0<p\leq 1$, that is, has its scale set by the natural scale in the problem, that of the BH radius\cite{SGOW,GWT,NVU}.   Such a scaling has the virtue that the spatial variation implied by the falloff of $G_{Ab}$ becomes weaker for larger BHs, in accord with the expectation that the near-horizon region of a very large BH is increasingly well-approximated by flat space.  The simplest possibility is $p=1$, though for example \cite{BoPe} have more recently argued for a new physics scale with $p=1/2$ based on arguments involving ``island" contributions to certain extensions of the integral over geometries (see further discussion in \ref{RWHS}).

Within such a limited domain for the $G_{Ab}(x)$, they in principle could also vary more or less rapidly with distance.  Again, if the scale of such variations were set to be microscopic, {\it e.g.} the Planck scale, then infalling observers would experience interactions with large momentum transfers, again yielding a large departure from the equivalence principle, and something that might be called a firewall.\footnote{Microscopic structure as with  the fuzzball picture of BHs is likewise expected to generically yield hard interactions with large momentum transfers\cite{SGOW}, and thus large departure from the equivalence principle.}  We instead assume a variation scale\cite{GWT,NVU} 
\beq\label{Ldef}
L=R^q
\eeq
that also grows with $R$, with $0<q\leq p$.  The simplest possibility is again $q=1$; notice also that if $p=q=1$, so that both scales are set to $\sim R$, that greatly limits the number of possible functions $G_{Ab}(x)$

These combined assumptions on $p$ and $q$, which can be described in terms of minimal  departure from the equivalence principle, align with our general approach of seeking minimal departures from the known physics of LQFT.  This can be thought of as a principle of ``nonviolence" of the interactions, for example for observers falling into large black holes. They can also be viewed\cite{BHQU} as part of a ``correspondence principle" for gravity, that in weak gravity limits, the fundamental theory is well approximated by LQFT plus semiclassical gravity.

Another aspect of this principle of nonviolence is that the interactions in \eqref{DeltaHI} should not significantly couple BH states with large energy differences, $\Delta E \gtrsim 1/R$; doing so would also result in a large departure from the spectrum predicted by Hawking\cite{Hawkrad}.  In fact we have already implemented this in the above description, by assuming the couplings are to states in the band between $M\pm\Delta M$.  A priori we can trivially extend the description to a wider range of states, with couplings to states with significantly larger or smaller energy than $M$, but we limit the range as part of the nonviolence assumption. 

\subsection{Universality}
\label{Univers}

A next question in our principled parametrization of ignorance is that of which operators $\calo^b$ are coupled by the interactions.  Here there are several possible guides.  First, the leading order Hawking process is approximately thermal, and fits well with the beautiful story of BH thermodynamics.  One way to understand this is as arising from the universal coupling of gravity to all fields, through the stress tensor.\footnote{To include gravitons in the Hawking radiation, one may augment this with their pseudo stress tensor.}  If, for example, the couplings \eqref{DeltaHI} only involved certain fields, that presents problems for detailed balance, if one tries to bring a BH into equilibrium with a thermal bath.
A related argument comes from the process of BH {\it mining}\cite{UnWamine,LaMa,FrFu,Frol}.  Perhaps the simplest form of mining involves threading a BH with a cosmic string.  Then, there are modes that propagate along the string, and the contribution of these modes speeds the evaporation process.  If we had interactions like \eqref{DeltaHI} that transferred information/entanglement to some modes, but not to the string modes, then the latter would lead to additional buildup of entanglement between the BH and the environment.  This would return us to the situation where the BH reaches the Planck mass with a large amount of remaining entanglement with the environment, necessitating information loss (violation of unitarity) or long-lived remnants.  This can be avoided by assuming a principle of universality, that the interactions \eqref{DeltaHI} responsible for the entanglement transfer couple universally to {\it all} modes.

There is a question about how to implement such universality.  Possibly, essentially universal behavior could be achieved by a collection of couplings such as \eqref{DeltaHI} to different operators involving all possible fields, although it is not entirely clear how to do so in a fashion which yields universal behavior.  But, we do have an operator that is universal -- the stress tensor $T_{\mu\nu}$, which couples to all fields.\footnote{Here, again, to include graviton interactions this can be extended to include a gravitational stress pseudo tensor.}  If we assume that the coupling goes through this universal channel, then \eqref{DeltaHI} becomes\cite{GiSh2,NVNLT,NVU}
\beq\label{DeltaHImet}
\Delta H_I = \int dV_x \sum_A \lambda^A G_A^{\mu\nu}(x) T_{\mu\nu}(x)\ 
\eeq
where now the more limited set of functions $ G_A^{\mu\nu}(x)$ parametrize our ignorance.  This coupling has an alternate interpretation: it behaves like a metric perturbation
\beq\label{Hdef}
H^{\mu\nu}(x) = \sum _A  \lambda^A G_A^{\mu\nu}(x)
\eeq
which is an operator depending on the quantum state of the BH.  In accord with the preceding discussion of correspondence and nonviolence, these perturbations extend over a range $\Delta R_a$  as in \eqref{Radef} from the semiclassical horizon, and have a spatial variation scale $L$ as in \eqref{Ldef}.  These scales limit the number of possible functions entering \eqref{Hdef}, particularly for $\Delta R_a\sim L\sim R$.  

\subsection{Interaction strengths}
\label{intstr}

Another important question is that of the size of the couplings, {\it e.g.} given by the typical size of the dimensionless $H^{\mu\nu}$.  This is constrained by the requirement that $\Delta H$ leads to ultimate unitarity.  The Hawking process builds up entanglement between the BH and environment at a characteristic rate
\beq
\frac{d S_{vN}}{dt} \sim \frac{1}{R}, 
\eeq
which yields a missing information $\sim S_{BH}$ by the time the BH evaporates to near the Planck mass.  The interaction $\Delta H_I$ must ``undo" this entanglement, and if it is to do so without a greatly different spectrum of emission, must therefore transfer entanglement from BH to environment at a rate 
\beq\label{inforate}
\frac{dI}{dt}\sim 1/R . 
\eeq
 This raises an interesting general question: if we have two subsystems which are coupled through a bilinear interaction, like our $\Delta H_I$ of \eqref{DeltaHI}, \eqref{DeltaHImet}, how fast does that transfer entanglement between the subsystems, given certain assumptions also about dynamics (such as thermalization or information transport) of the individual subsystems?  

We first note that if $|\tilde\psi\rangle$ is a typical BH state, and if the typical size of $H_{\mu\nu}$ is given by
\beq\label{strongnvu}
\langle\tilde\psi|H_{\mu\nu}(x) |\tilde\psi\rangle \sim \calo(1)
\eeq
 then that seems to easily transfer information at the required rate
\eqref{inforate}.  The reason is that this means the coupling is behaving like a classical $\calo(1)$ metric fluctuation which is varying on spatial and temporal scales $\sim R$; this can easily produce quanta at rates $\sim 1/R$, which carry the outgoing information.

But, in the spirit of minimality, there is a question of the {\it necessary} size.  Ref.~\cite{NVU, GiRo} began a study of the general rate question, using {\it e.g.} toy models, but a simple heuristic is that if a coupling between subsystems can induce transitions at a given rate -- an example would be an interaction causing an atom to decay and emit an outgoing photon -- then that also gives an estimate of the rate at which the interaction can transfer information.  While couplings of size \eqref{strongnvu} are sufficient to transfer information at the needed rate, it appears that this can also be done with weaker interactions; for this reason the size \eqref{strongnvu}  will be referred to as corresponding to a {\it strong} version of nonviolent unitarization.

To estimate the interaction size of a minimal, {\it weak}, version of nonviolent unitarization, following the preceding heuristic, we need to estimate the rate at which $\Delta H_I$ induces transitions.  
A simple estimate of that rate comes from first-order perturbation theory, via Fermi's Golden Rule, 
\beq\label{FermiGR}
\frac{dI}{dt}\sim \frac{dP}{dt} = 2\pi \rho(E_f) |\Delta H_I|^2\ ,
\eeq
where $\rho(E_f)$ is the density of final states of the system, and $|\Delta H_I|$ is the typical size of a matrix element of $\Delta H_I$.  
At this point we see that there is apparently a way to achieve the rate \eqref{inforate} with much smaller couplings: for generic couplings between states the density of final states includes the large factor $N\sim \exp\{S_{bh}\}$ of the number of BH states, and so very tiny couplings, of typical interaction size
\beq\label{weaknvu}
 |\Delta H_I|\sim \frac{1}{\sqrt N} = e^{-S_{bh}/2}
 \eeq
 appear to suffice.\footnote{A similar numerology, but corresponding to a different proposed physical mechanism, was described by Mathur\cite{MathureS}.  In Mathur's case, the small amplitude describes transitions to fuzzball states from a BH, and the large number of such fuzzball states leads to an analogous compensation.}

\subsection{Summary of principles and parametrization, and questions}

To summarize this section, adopting a  conservative ``minimalistic" approach, we have begun a principled parametrization of ignorance,
investigating the simplest interactions that can undo the problematic entanglement or missing information generated by the Hawking process, and restore ultimate unitarity of the BH decay process.  In general, this involves couplings \eqref{DeltaHI} between the BH and environment subsystems.  

A correspondence principle to LQFT and semiclassical gravity implies that these couplings only extend a distance $\Delta R_a$ from the semiclassical horizon, over a region which we will refer to as a {\it quantum halo} of the BH, and moreover that their spatial and temporal variation is on scales that grow with $R$, as in \eqref{Ldef}.  This notion of nonviolence can alternately be thought of as part of seeking a minimal violation of the equivalence principle.

The desire to at least approximately preserve BH thermodynamics, the universal nature of gravity, and Gedanken experiments involving mining all point to the hypothesis that the couplings are universal.  A simple way to achieve this  that is via a coupling to the stress tensor, of the form \eqref{DeltaHImet}.  Such couplings can be thought of as describing BH state-dependent metric perturbations.

The necessity of ultimate unitarity implies that the couplings must have a sufficient strength.  While the strong scenario, with \eqref{strongnvu}, suffices, apparently one can achieve the necessary information transfer rate with couplings that are in a sense much smaller, with typical matrix elements given by \eqref{weaknvu}.  

The resulting limitations are significantly constraining.  The simplest interactions are parametrized by the finite number of functions $G_A^{\mu\nu}(x)$.  The limited support of these functions, to the range $\Delta R_a$ from the horizon, combined with the limit $L$ on their spatial variation scale, greatly limits the number of possible independent functions; for example, for $\Delta R_a\sim L\sim R$, this number is $\calo(1)$ rather than very large.  These serve as a starting point for a more explicit parameterization; see \cite{GiPs} for some initial work on this.  Beyond giving such systematic parametrizations, and seeking further constraints on the couplings, other important questions remain.

The first is what underlying fundamental quantum gravity dynamics explains such couplings and determines their values.  While comments were made on this above, and this will also be touched on in section \ref{qgfound}, we view this as a very important open question for future work, and stress that the parametrization given in this section is to be viewed as a parametrization of an effective description of the important effects of the more basic physics.  Of course, ultimately such effective descriptions may well give very important clues regarding the structure of the more fundamental dynamics.

Another question is whether there are any inconsistencies in such an effective theory, either internal or with known aspects of physics, or if there are other compelling objections, for example as a result of general principles.  We will briefly discuss some criticisms that have been raised in the next section.

Finally, corrections to the evolution that extend outside the would-be horizon imply that matter and fields near the BH can interact differently with it than with a classical BH.  The revolutionary opening of observational windows on the near-horizon regime, both through very long baseline interferometry and through observation of gravitational waves, raises the very important question regarding whether such new interactions could in fact have observable effects.  We will briefly discuss some work on this question in Sec.~\ref{sigs}.

\section{Responses to some criticisms}
\label{response}

This section will briefly discuss some of the common prejudices, concerns, and objections regarding the basic ideas of nonviolent unitarization, and will provide some responses to them.  Since it is an important question how to reconcile quantum BH evolution with general principles, it is in particular important to understand whether NVU encounters any essential conflict with known physics or general principles.  

\begin{enumerate}
\item{{\sl Shouldn't any new physics be restricted to the immediate vicinity of the singularity?} }  This has been a longstanding belief in much of the quantum gravity community, but the nature of the essential conflict of the unitarity crisis appears to show this cannot be correct.  For example, delaying new physics to the final stages of BH evaporation yields either information loss or microscopic remnants, whose problematic nature was reviewed above.  After decades of discussion, we appear to have learned that the unitarity crisis is demanding new physics in the vicinity of the would-be BH horizon -- and now most proposals on how to resolve the crisis involve some kind of new physics on these horizon scales.

\item{{\sl Shouldn't any new physics be restricted to within a planckian distance of the would-be horizon?} }   While this has been a common prejudice, it is not clear that there is any principled argument in its favor.  And, in fact, there are now multiple reasons to question this assumption.  First, such an assumption -- as discussed above -- generally leads to violent behavior near the horizon\cite{SGTrieste,BPZ,AMPS}, {\it e.g.} on infalling observers, aka ``firewalls."  Put differently, this assumption leads to a large violation of the equivalence principle.
Secondly, a basic motivation for this assumption, that the Hawking radiation is somehow ``produced" in a microscopic neighborhood of the horizon, appears to be a {\it basis dependent} statement without unambiguous foundation.  Reexamination of this argument from multiple perspectives argues that instead the Hawking radiation is produced in a quantum  ``thermal atmosphere" of thickness $\sim R$ around the BH horizon\cite{SGstefanB,SEHS,SE2d,GiPe1}.  Third, restricting new physics or interactions to an $\calo(l_{Pl})$ vicinity appears to require an unnatural fine tuning; there is a natural scale in the problem -- given by the horizon size, or by that of the thermal atmosphere, so it seems most natural for new physics to arise on some longer scale set by that.  Finally, there are now even arguments from other approaches\cite{BoPe} for the same kind of nonviolent scaling, with $\Delta R_a\sim R^{1/2}$ as in \eqref{Radef}.

\item{{\sl New interactions, such as described in \eqref{DeltaHI}, clearly violate the Equivalence Principle.}}  The Equivalence Principle is a key principle underlying {\it classical} general relativity.  However, without classical spacetime and classical geometry -- which appear to be approximations -- it is not clear how to formulate it.  Indeed, one appears to see departures from it even from leading quantum effects, such as QFT effects related to Hawking radiation.  Perhaps we should not be surprised if the complete quantum theory, which goes beyond classical geometry, involves some modification to the Equivalence Principle.  What does seem surprising is an extreme violation, such as would be associated with interactions with a microscopic scale $\Delta R_a$, producing a firewall.  The nonviolent scaling \eqref{Radef} yields a much more mild departure from the classical Equivalence Principle.

\item{{\sl If one considers BH mining gedanken experiments\cite{UnWamine,LaMa,FrFu,Frol}, the new interactions no longer lead to unitary evolution.}}  Indeed, \cite{AMPSS} went so far as to say that unitarity would require ``an implausible conspiracy" with the mining equipment.  However, while further careful examination of mining arguments are merited, as was argued above, universal couplings to matter/fields appears to largely address the objection.  Mining -- {\it e.g.} using a cosmic string to increase the number of excitations in the Hawking radiation -- increases the rate at which a BH shrinks, yielding a possible inconsistency.  But, if the information-transferring couplings are universal, then the information transfer from the BH will also increase in a commensurate fashion.

\item{{\sl The interactions of NVU present an inconsistency with BH thermodynamics}}.  Again, this appears to be to a significant degree addressed if the couplings are universal.  However, this question also merits further careful investigation.  For example, the interactions of \eqref{DeltaHI}  are expected to increase the rate at which the BH emits radiation.  However, since the BH is a localized object, if it thermally emitted the rate would be given by the Stefan-Boltzmann law.  A greater emission rate may be interpreted as due to the change of the effective size of the emitting surface, rather than due to a change of the temperature-energy relationship.  It is true that in the presence of such interactions the formula for BH entropy $S_{bh}$ may not necessarily yield the exact Bekenstein-Hawking entropy $S_{BH}$; some discussion of this appears in \cite{SMBH}.  However, the status of the Bekenstein-Hawking entropy as the true entropy of the BH is also something that is not well-understood from a microscopic perspective. 

\item{{\sl The interactions \eqref{DeltaHI}  are acausal}}.  It is true that, in Minkowski space, nonlocal interactions can yield acausal signaling.  The reason is that an interaction transferring information at spacelike separation transfers information backward in time (but still at spacelike separation) from the viewpoint of a boosted frame.  Two such interactions can be combined to transfer information into the backward lightcone of an observer, yielding acausality.  However, if the interactions \eqref{DeltaHI}  are associated to the structure of the quantum BH, and localized in its vicinity, it is not clear there is an analogous argument.  Specifically, the BH classically specifies a frame, and the symmetries are different; if, for example, the information transfer is associated with a time delay in that frame, it is not clear how any observer experiences an inconsistency\cite{NLvC}.  Of course, part of understanding the more fundamental description of the theory is understanding why such interactions are present in the vicinity of BHs but not elsewhere -- but that is beyond the scope of our effective description.\footnote{It might also be noted that such bilocal interactions have subsequently been increasingly investigated in the context of BHs in AdS/CFT, beginning with
\cite{Maldacena:2017axo}.} 

\item{{\sl The interactions \eqref{DeltaHI}  are slicing dependent.}  In the absence of a more fundamental understanding of gauge invariance in full quantum gravity, the structure of the effective theory has been described after choosing a classical analog of a gauge, namely a slicing.  The question of full gauge invariance remains one for a deeper understanding of the more complete theory.}

\item{{\sl It is not clear that AdS/CFT predicts such interactions.}}  The question of {\it how} AdS/CFT solves the unitarity crisis has remained a central one since it was first proposed.  A significant part of that problem is finding the ``holographic map" which allows one to describe quantities in the ``bulk" theory of gravity in AdS in terms of the boundary quantities of the CFT.  Indeed, one of the most convincing proposals for explaining the holographic map\cite{Maro1,Maro2,Jaco,JaNg,HoloUn}, as noted above, arises from the assumption that the gravitational constraints have been solved.  But, as pointed out in \cite{HoloUn}, that would mean that in order to construct the holographic map, we must first have a description of evolution in the bulk -- an unsatisfying circularity.  That in particular seems to mean we would have to understand BH evolution in order to construct the holographic map, rather than determining BH evolution from the boundary theory plus a holographic map constructed by other means. Since AdS/CFT is the leading approach to addressing the crisis from string theory, it is also not clear that there is anything else in string theory that will resolve the problem.

\end{enumerate}

\section{Possible observational signatures}
\label{sigs}

Our discussion of the ``black hole theorem" suggests that to avoid inconsistency we must have some interactions that make the evolution of the BH exterior dependent on the BH quantum state -- these interactions extend outside the semiclassical horizon.  Indeed, while there is not yet consensus in the community, many of the current proposals for a resolution of the unitarity crisis share this feature, that there are new interactions or structure outside what would be the BH horizon.  Since we are now able to observe the vicinity of the BH horizon through very long baseline interferometry\cite{EHT} and through gravitational wave detection\cite{LIGO}, that suggests the possibility of {\it observational} effects of any such new physics.

\subsection{Gravitational wave signatures}

For example, new interactions or structure outside the horizon would be expected to modify the scattering of gravitational waves (GWs) from that of a classical BH.  When two BHs inspiral into each other, a component of the GW signal arises from the scattering from the individual BHs; one can think of one BH as emitting GWs as it orbits, and some of those waves scatter from the other BH and contribute to the outgoing GW signal.

One can therefore adopt, as an intermediate parameterization, the amplitudes for scattering of GWs from a BH, and investigate the effect of their deviation from classical expectations.  That approach has been recently taken in \cite{GiFr}, building on earlier work by \cite{Hughesetal,MVP} and others.  The effect on GW observations is most easily studied, as in \cite{GiFr}, in the case of an extremal mass ratio inspiral (EMRI), which is well-approximated as a one-body problem.  One can infer the contribution of the scattering amplitude modification $\Delta \calt$ to the outgoing signal by Green function methods.  Of course, for small $\Delta \calt$ the modification is small, and may be challenging to detect directly in the observed GW signal.  But, there is also an amplifier: $\Delta \calt$ can change the {\it rate} of energy emission from the inspiral, and if the latter has many orbits, then one can build up  a significant net phase shift, which is potentially more easily observed.  For example, initial estimates\cite{GiFr}, extending\cite{Hughesetal}, suggest possible sensitivity to a benchmark scenario for near horizon interactions, parameterized in terms of a near-horizon reflection coefficient, with sensitivity to such coefficients 
of size down to $\calr\sim 10^{-6}$.

It is generally expected that more extreme scenarios, such as massive remnants (gravastars, fuzzballs, firewalls, frozen stars, \ldots) would have substantial $\Delta \calt$, perhaps making them easier to rule out, but our focus is on NVU.  In particular, the weak scenario for NVU is most conservative, and seems most difficult to probe.  But, simple estimates suggest the possibility of doing so, despite the ``smallness" of the interactions.  To see this, consider scattering a graviton from a quantum BH which has an effective description including the interactions $\Delta H_I$ of \eqref{DeltaHImet}, including a gravitational pseudostress tensor.  The probability for the interaction to cause a transition in the graviton state can also be estimated by Fermi's Golden Rule, \eqref{FermiGR}, with a rate
\beq\label{FGRG}
\frac{dP}{dt}=2\pi \rho(E_f)|\langle\beta|\Delta H_I|\alpha\rangle|^2
\eeq
where now the matrix element is between an initial graviton state together with the initial BH state, together denoted $|\alpha\rangle$, and a final state with a scattered graviton and another BH state, $|\beta\rangle$.  
In the weak scenario where typical couplings are exponentially small as in \eqref{weaknvu}, one can still have a significant ({\it e.g.} $\calo(1/R)$) transition rate as a result of the exponentially-large density of final states.  This suggests that the interactions can yield $\Delta \calt = \calo(1)$ at the quantum level. 

Scattering of this size seems well within the possible sensitivity described above.  One additional caveat is that the classical $\Delta \calt$ arises from a large number of gravitons scattering from a quantum BH, and this must be related to the quantum $\Delta \calt$ whose calculation we have just outlined.  A more precise connection between these scattering amplitudes is a subject for future work.

Of course, the strong scenario of NVU is expected to directly yield a significant $\Delta \calt$, and thus to plausibly be well within sensitivity of overall phase measurements.  One can also consider other effects, including the  direct contribution to the outgoing GW signal\cite{GiFr}.  Most recently, \cite{SeCh} have considered additional stochastic contributions to GW signals that could originate from this strong scenario, and that possibly fall within range of measurement through future terrestrial-based GW observations.

\subsection{Electromagnetic signatures}

Very long baseline interferometry (VLBI) has allowed us to observe light from the near-horizon region of supermassive BHs, as was first spectacularly demonstrated by the Event Horizon Telescope (EHT)\cite{EHT}.  As with the preceding discussion of scattering of GWs, we can investigate the effect of, {\it e.g.}, the universal interactions of \eqref{DeltaHImet} on light propagation near the horizon.  Once again, at the level of a single photon, the interactions yield estimated scattering probabilities given by the Fermi's Golden Rule formula \eqref{FGRG}, now with photon states replacing graviton states.  Again, the compensation of large and small exponentials apparently can produce an $\calo(1)$ probability.  However, now there is an important difference -- for nonviolent interactions, the scaling of section \ref{scaless} tells us that the $G_A^{\mu\nu}$ of \eqref{DeltaHImet} have typical variation on distance scales $\sim R$, in the simplest case, and therefore that the $\calo(1)$ transition amplitudes are between states whose momentum differs by $\sim 1/R$.  Since the radiation observed by EHT is in the mm to sub-mm range, and we are observing supermassive BHs with radii $\sim 10^7-10^{10}$ km, even such $\calo(1)$ scattering at first sight appears to have a negligible effect on the signal.\footnote{The same kind of argument can be made for any other short wavelength matter or radiation scattering from the BH, indicating tiny momentum transfers and thus very small effect, and also extended to scattering of matter which enters the BH -- specifically to infalling observers.}

In the event a strong NVU scenario is the correct description, there can be much more significant effects on observation.  The interactions of strong NVU mimic significant classical fluctuations in the metric, on spatial and temporal scales $\calo(R)$.  A preliminary investigation of the effect of such fluctuations on images seen through VLBI was performed in \cite{GiPs}.  The effective fluctuations can produce quite dramatic time dependence of the VLBI images, and even significant distortions of individual snapshots.  These are significant enough that it appears that with further analysis present observations can begin to rule out a range of parameters of these scenarios, even in the absence of continuous observation over the relevant timescales, which are of order months for M87. Needless to say, observation of any  image distortion fitting these scenarios would be very important to explore.

\section{Implications for quantum gravity foundations}
\label{qgfound}

Our discussion of the ``black hole theorem" in Sec.~\ref{bhthms}, and of the treatment of LQFT and perturbative gravity in Sec.~\ref{LQFTQGR}, has attempted to lay out a succinct summary for why the crisis is such a crisis, calling for a radical departure from the known physics of LQFT and semiclassical and perturbative GR.  The possible resolutions are highly constrained by general principles, and most na\"\i ve proposals either seem to run into inconsistencies -- either internally or with such general principles -- or to need a large amount of additional structure, raising additional questions and problems.  In short, the unitarity crisis appears to imply a failure of LQFT and semiclassical/perturbative GR to describe evolution of a BH in our quantum universe, and is strongly indicating the need for some modification of the principles underlying the foundations of known physics.  The need to resolve the crisis thus seems like it will play a key role in the foundations of quantum gravity, and the nature of the problem is plausibly providing important clues towards those foundations.

One important message seems to be that quantum gravity modifications to our known physics are not just restricted to Planck scales, as has been widely assumed in much of the discussion of quantum gravity not aimed at understanding quantum BHs.  A systematic outline of consequences of possible modifications that are restricted to Planck scales appears to result in the unacceptable and unphysical alternatives described in Sec.~\ref{bhthm}.  Resolution of the crisis appears to demand new physics on scale sizes of order the size of whatever BH is being studied, which can in principle be enormously large.  This is evidenced not only by general arguments, but also by the near-convergence of attempted resolutions on {\it some} new physics on these scales.

Needless to say, important modifications to such long-scale physics would be both radical, and are expected to be highly constrained by our body of observed physics.  It is difficult to provide scenarios without significant additional structure -- {\it e.g.} associated with massive remnant scenarios such as fuzzballs, gravastars, firewalls, {\it etc.}, resulting in new objections and potentially problematic behavior.  Instead, in NVU one takes the viewpoint that one should begin by parameterizing the effects needed for consistency, with a minimum of extra structure and minimal deviation from known physics.  Difficulties for extra structure are one motivation for this.  But there is another important one, arising from the fact that time and again physics has taken the path of minimal -- but profound -- modification of basic structure when a theory with new principles replaces a previous physical theory.

Of course ones understanding is not complete until one has a deeper understanding of the principles of a more basic explanation of the underlying physics.  As an effective description, NVU is not directly addressing that -- but one expects that if it captures the correct effective physics, it {\it is} providing very important clues toward the nature of that more complete foundation.  It is certainly worth considering what that more basic physics could look like, in working toward its ultimate formulation.

The remainder of this piece will briefly discuss two possibilities for a deeper explanation of the interactions of NVU.  One is based on an extrapolation of the geometric constructions of GR, involving spacetime wormholes and baby universes.  A second is based on the possibility that spacetime is not fundamental, but instead emerges from a more deeper quantum reality.  (These may, possibly, even overlap.)

\subsection{Replica wormholes and baby universes}
\label{RWHS}

The dynamical geometry of GR suggests extrapolation to dynamical topology, which has long been considered from various viewpoints.  A significant fraction of the community has become excited that a  set of hypothesized rules for extending the euclidean sum over geometries yields entropy formulas that match the rise of the BH entropy with the entanglement growth exemplified by  \eqref{qbitevol} in the initial stages of BH evaporation, but then transition to a decreasing entropy as is expected for a shrinking system, as was concretely spelled out by Page in \cite{Page1,Page2}.

These rules are based on formal prescriptions for summing over ``replica geometries," which have nontrivial topology evidenced through ``replica wormholes;" for a recent review see \cite{RWHrev}, and for some discussion of underlying assumptions see\cite{GiTu}.  They do not yet give a set of quantum amplitudes between states -- say described in the setting of lorentzian evolution -- from which the entropy formulas could be computed from a more basic viewpoint.  However, it seems that if there is a more basic explanation of the physical mechanism at work, that arises from a more complete description of spacetime topology change, and specifically from the emission and absorption of baby universes, as was emphasized by Marolf and Maxfield\cite{MaMa}.

Such baby universe emission/absorption began to be seriously considered in the 1980s\cite{LRT,HawkingBU,GiSt1}.  While it was initially believed to lead to information loss from a large-scale ``parent universe" such as our own\cite{LRT,HawkingBU,GiSt1}, a more systematic investigation showed that, at least  in a certain approximation, this is not true\cite{Cole,GiSt2,GiSt3Q}.  Specifically, the emission of a baby universe (BU) from our universe can be parameterized by an interaction of the form
\beq\label{buints}
H_{BU}= a_i^\dagger \calo_{BUi} +h.c.\ ,
\eeq
where $a^\dagger_i$ acts on the state of BUs to create an additional BU of type $i$, and $\calo_{BU,i}$ is an operator that parameterizes the effect of that emission on the fields in our Universe.  This ``third-quantized" hamiltonian, which is analogous to those for particle emission from a system in traditional second quantization, may also need to be supplemented by terms describing BU interactions or other effects\cite{GiSt3Q}, which will be neglected in the present approximation.  It has been proposed\cite{MaMa} that such interactions with an ensemble of BUs could explain the underlying physics behind the replica wormhole calculations, and in particular the decrease of entropy of an evaporating BH.

In examining this claim, one can go back to the older discussion of \cite{Cole,GiSt2}.  There it was argued that one can find BU states that diagonalize the operators $a_i+a_i^\dagger$, with eigenvalues $\alpha_i$, and that in these ``alpha eigenstates" the effects of the interactions \eqref{buints} are simply new terms in a standard hamiltonian or lagrangian, with coupling constants given by those eigenvalues.  This relied on the operators $\calo_{BUi}$ not carrying net energy or momentum (or other local charges), as would be expected when emitting a separate closed universe.  Also part of this picture is that the operators $\calo_{BUi}$ will be integrals of operators that are local on scales large compared to the associated BU size $R_i$, and so in the long distance effective theory one simply has a modification to various coupling constants of the local effective  theory.  Further details are explained in the preceding references, and in the large body of literature that followed.

In such a picture, one could ask how BUs could contribute to unitarization of BH evolution.  One possibility is that BU interactions are important, in which case the preceding discussion needs to be modified, but let us assume that we work in the noninteracting BU approximation.  Then, if we work in an alpha eigenstate, physics in our universe is described by an ordinary LQFT hamiltonian or lagrangian, together with extra contributions of operators $\calo_{BU,i}$ with couplings given by the $\alpha_i$'s.  

On distance scales large compared to the $R_i$'s, this is once again an effective LQFT, and so one seems to be returned to the discussion of section \ref{bhthm}, in which the crisis remains.
What is new, now, is that the replica wormholes can have scales comparable to the BHs whose evolution they modify, and one expects this to be true of the corresponding BUs, although a more concrete description of the emission processes and corresponding operators is still needed.  But this suggests that in \eqref{buints} the operators in the hamiltonian can be nonlocal on the BH size scale $\sim R$.  And, it appears that can make contact with the more general NVU discussion involving \eqref{Heffect} and \eqref{Hsep}, and provide a mechanism (BU emission) for the interactions of NVU.

This is not yet a complete story, and leaves several questions.  The first regards the actual form of the hamiltonian \eqref{buints} and its operators, and how this yields unitary evolution.  First, the couplings depend on the alpha eigenstate of the BUs, and so in principle could be weak or strong.  As was outlined in Sec.~\ref{intstr}, one seems to need a specific size interaction strength to achieve the needed information transfer rate for ultimate unitarity.  Secondly, we do not have a good understanding of which operators enter \eqref{buints}.  One associated question is how such a collection of operators achieves universality, as described in Sec.~\ref{Univers} -- if that is indeed the correct physics, and if not, what is the alternative.  

One other concern is that such a scenario is tied to a traditional geometrical picture of spacetime, which from other viewpoints we expect to not necessarily be fundamental.  An interesting point, if this {\it is} the correct scenario, is that it seems to argue that topology changing processes are {\it required} in order for the consistency of quantum gravity with quantum unitarity.

\subsection{An origin in a foundational picture for quantum spacetime}

There are ample reasons to believe that spacetime may not be fundamental. One is that it is very hard to make sense of a spacetime description at distances shorter than the Planck length both from the point of view of theory, or from the point of view of any experiment, Gedanken or otherwise, to probe this regime.  And, if manifold structure is not part of the fundamental description at short distances, it is apparently only an approximate description of the structure at any distances.

If this is true, an important question is what {\it is} the mathematical starting point for a description of physics.  Since quantum mechanics is both well-tested and appears difficult to modify, an attractive possibility is that the correct fundamental structure is that of quantum mechanics, and is based on Hilbert space or an appropriate generalization, with additional mathematical structure needed to describe the physics of gravity and other excitations\cite{QFG,QGQF}.  This can be motivated from the algebraic description of LQFT (see {\it e.g.} \cite{Haag}), in which the basic structure is Hilbert space together with a net of local algebras that encode the structure of the underlying manifold.

Indeed, this discussion connects to another possible objection to NVU, regarding the lack of an obvious mechanical picture behind the interactions of NVU, if they do not for example arise from baby universe emission as described above.  However, there are different possible perspectives on this.  As an example, we commonly think of particles and their interactions as providing a mechanical picture for much of physics.  But, a different perspective is that one can start with the principles of quantum mechanics, relativity, and locality and {\it derive} LQFT as a theory reconciling these principles, in the process discovering the origin of particles from the more basic set of principles.\footnote{Examples of this basic approach are provided in \cite{Wein,ColeQFT} as well as in the algebraic approach to the subject.}  This viewpoint also illustrates the power of effective quantum theories, since one can begin with the problem of formulating a quantum-mechanical theory respecting these further principles, and arrive at LQFT.  From such a quantum-first perspective, the more fundamental question is what principles, beyond those of quantum mechanics, furnish the correct starting point.  

If spacetime is not fundamental and a more basic quantum construct is, then a description of physics in terms of local quantum fields on spacetime is necessarily only an approximation, and is expected to be subject to corrections.  Possibly, the interactions of NVU are part of the corrections to an erroneous description of physics in terms of fundamental geometrical spacetime, and so parameterize some of the effects of the departure of that description from 
the true theory.  If so, then better understanding of these interactions is likely to provide us with important clues to the more fundamental quantum dynamics underlying gravity and lying at the foundations of physics.

\begin{acknowledgement}
I thank my multiple collaborators on these subjects over the years, and also numerous colleagues for many discussions.  I also thank R. Bousso and G. 't Hooft for comments on an earlier version.  
The current work is supported in part by the Heising-Simons Foundation under grants \#2021-2819, \#2024-5307, and by the U.S. Department of Energy, Office of Science, under Award Number DE-SC0011702.
\end{acknowledgement}

%%%%%%%%%%%%%%%%%%%%%%%%%%%%%%%%%%%%%%%%%%%%%%%%%%%%%%%%%
%
%
%
%
%
%\section*{Appendix}
%\addcontentsline{toc}{section}{Appendix}
%
%When placed at the end of a chapter or contribution (as opposed to at the end of the book), the numbering of tables, figures, and equations in the appendix section continues on from that in the main text. Hence please \textit{do not} use the \verb|appendix| command when writing an appendix at the end of your chapter or contribution. If there is only one the appendix is designated ``Appendix'', or ``Appendix 1'', or ``Appendix 2'', etc. if there is more than one.
%
%
%
%
%
%%%%%%%%%%%%%%%%%%%%%%%%%%%%%%%%%%%%%%%%%%%%%%%%%%%%%%%%%%
%
%
%
%
%
%\biblstarthook{References should be \textit{cited} in the text by numbers.\footnote{Please make sure that all references from the list are cited in the text. Those not cited should be moved to a separate \textit{Further Reading} section.} The reference list should be \textit{sorted} in alphabetical order. If there are several works by the same author, the following order should be used: 
%\begin{enumerate}
%\item all works by the author alone, ordered chronologically by year of publication
%\item all works by the author with a coauthor, ordered alphabetically by coauthor
%\item all works by the author with several coauthors, ordered chronologically by year of publication.
%\end{enumerate}
%For the reference style, we suggest to use \textit{LaTeX (US)} from INSPIRE.}

%%%%%%%%%%%%%%%%%%%%%%%%%%%%%%%%%%%%%%%%%%%%%%%%%%%%%%%%%

\end{document}